# PRODUCTION OF HIGH-$P_T$ CUMULATIVE PARTICLES IN PROTON-NUCLEUS INTERACTIONS AT 50 GeV.

**V.V.Ammosov[1], N.N.Antonov[1], V.A.Viktorov[1], V.A.Gapienko[1], G.S.Gapienko[1], V.N.Gres'[1], V.A.Korotkov[1], A.I.Mysnik[1], A.F.Prudkoglyad[1], Yu.M.Sviridov[1], A.A.Semak[1], V.I.Terekhov[1], V.Ya.Uglekov[1], M.N.Ukhanov[1,\*], B.V.Chuiko[1],**

**A.A.Baldin[2,3], S.S.Shimanskiy[2,\*\*]**

[1]*Institute for High Energy Physics, Protvino, Russia*

[2]*Joint Institute for Nuclear Research, Dubna, Russia*

[3]*Institute for Advanced Studies "OMEGA", Dubna, Russia*

E-mail: [\*] Mikhail.Ukhanov@ihep.ru, [\*\*] Stepan.Shimanskiy@jinr.ru

The data on production of cumulative particles in the high transverse momenta domain (up to ~ 3.5 GeV/c) in proton-nucleus interactions are presented for the first time. An indication on the local character of particle production in the cumulative domain is obtained. The observed strong dependence of the particle production cross section on the atomic mass of the target does not fit the A - dependence obtained in the pre-cumulative and cumulative domains for low transverse momenta in the constant part of the exponent. The experiment was performed at U70 (IHEP) with the extracted 50 GeV/c proton beam.

## INTRODUCTION

The study of cumulative processes, i.e., processes in the kinematic domain forbidden for scattering on free nucleons, showed (see, e.g., [1]) the multinucleon (multiquark) configurations with a size comparable with that of the nucleon can probably exist in nuclear matter. The nature of these configurations has not been completely elucidated. Most experimental data on cumulative processes were obtained in the form of inclusive spectra in the nucleus fragmentation domain of a target or a projectile and with low transverse momenta. In the kinematic domains mentioned above, it is difficult to study other particles participating in cumulative particle production, since the recoil particles of a multinucleon (multiquark) configuration can hardly be kinematically identified, among other nuclear fragmentation products.

The study of cumulative processes in the high domains makes it possible to detect the cumulative particle that can be used as a trigger for the interaction with a multinucleon configuration, and can kinematically to separate the recoil and the nuclear fragmentation products. Theoretical estimates [2] show that in $x_T = \frac{p_T}{p_T^{\max}} \approx 1$ domain an interaction with multiquark configurations, rather than multiple interactions in a nucleus, are expected to be dominating.

An important feature of cumulative processes and $pA$ - production of non-cumulative particles with high $P_T$ momenta is an enhanced A-dependence of inclusive cross sections. If one parameterizes cross sections of these processes by a power function of the atomic mass number A of interacting nuclei, $A^\alpha$, the power α is higher than unity, which indicates the collective character of the mechanisms of production of such particles.

This paper is devoted to the experimental study of inclusive hadron production in the high-$P_T$ pre-cumulative and cumulative domains, investigation of A-dependences in these reactions and study possibility of the local interactions with multinucleon configurations. Such an experimental investigation was performed for the first time.

## EXPERIMENTAL TECHNIQUE

The data presented here were obtained with a single-arm magnetic spectrometer SPIN [3] (IHEP, Protvino), the schematic diagram of this spectrometer is shown in Fig.1. A 50 GeV/c proton beam with an intensity of ~5×10$^{12}$ ppp interacted with thin (0.6 - 0.8 g/cm$^2$) nuclear targets made of carbon, aluminum, copper, and tungsten. The system of seven magnetic elements makes it possible to select charged particles with a certain momentum and angle. The angular acceptance of the setup is Δφ≈100 mrad with respect to the azimuthal angle and Δθ≈40 mrad with respect to the polar angle. The momentum acceptance of the setup varies from 5.5% at 1 GeV/c to 3.5% at 6 GeV/c. The setup includes the trigger, the spectrometer and the time of flight systems. The experimental details are given in [3].

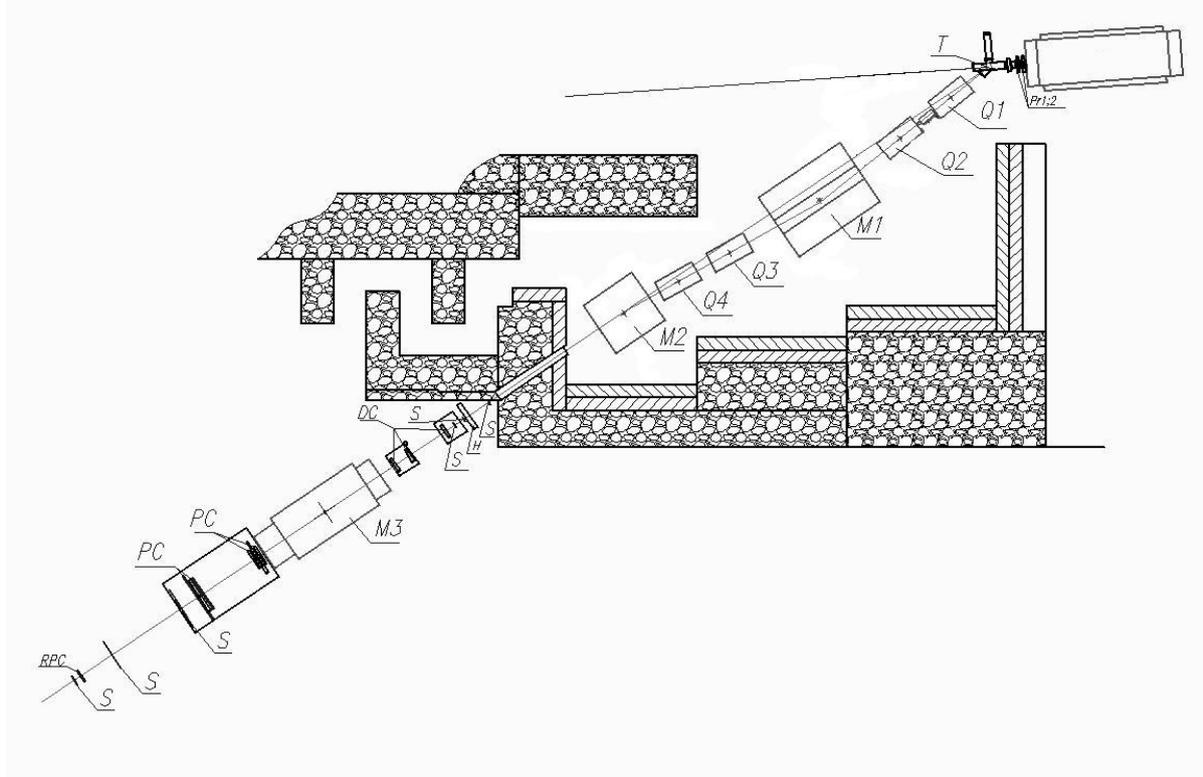

Fig. 1. Schematic diagram of the setup: T - target; Q1, Q2, Q3, and Q4 – quadrupole magnetic lenses; M1, M2 – dipole magnets selecting particles emerging from the target at different angles; M3-analyzing magnet; S - set of scintillation trigger counters; PC, DC -wire chambers of the tracking system; H (hodoscope), and RPC (resistive plate chamber) – elements of the time of flight system.

In this study charged particles dipole magnets selecting particles emerging the target at $35^0$ (lab. frame), were registered in a momentum range from 1 GeV/c to 6.6 GeV/c, which corresponds to the transverse momenta from 0.57 GeV/c to 3.76 GeV/c. The maximum momentum for negatively charged particles was 5.8 GeV/c. The momentum spectra for all four targets were obtained in the form of double differential cross section determined as follows:

$$\frac{d^2\sigma}{dP \cdot d\Omega} = \frac{A}{N_A \cdot \rho \cdot t} \cdot \frac{1}{\varepsilon \cdot \Delta P \cdot \Delta \Omega} \cdot \frac{N^h}{N_{prot}} \qquad (1)$$

where $A$ is the number of nucleons of the nucleus, $N_A$ is the Avogadro constant, $\rho$ is the target density, $t$ is the target thickness, $\Delta P$ and $\Delta \Omega$ are the setup momentum and solid angle acceptances, $N_{prot}$ is the total number of protons passed through the target, and $N^h$ is the number of registered particles at the given momentum. The quantity $\varepsilon$ takes into account the losses in the spectrometer arm and the efficiency of the trigger system. The quantity ε·ΔP·ΔΩ was calculated using GEANT3 [4] taking into account the parameters of all elements of the setup.

## EXPERIMENTAL DATA

Fig.2 shows the inclusive production cross sections for positively and negatively charged particles as functions of momentum. The upper axis shows the corresponding transverse momentum. Vertical dashed lines show kinematic thresholds for nucleon-nucleon interaction. According to the data from the time of flight system, for a momentum higher than 4 GeV/c the fraction of π- and K-mesons among positively charged particles is about 5%. Therefore, the kinematic threshold for positively charged particles is shown for elastic nucleon-nucleon scattering. For negatively charged particles the kinematic threshold of $\pi^-$-meson production in a free-nucleon interaction is shown. It can be seen from this figure that for high $P_T$ ($P_T > 2.5$ GeV/c) both positively and negatively charged cumulative particle production are observed.

The errors shown in Figs. 2-5 take into account only statistical uncertainty. The inaccuracy of determination of the number of beam particles passing the target gives the main contribution into the systematic error. The estimate of the systematic error in these cross sections was made by comparing the data obtained at different times. It is independent of momentum and can reach about 20%. Since different targets were irradiated in close conditions, the systematic error for the cross section ratio is smaller than that for the cross sections. The systematic error for the ratios was estimated as ~7% by comparing repeated measurements at different times.

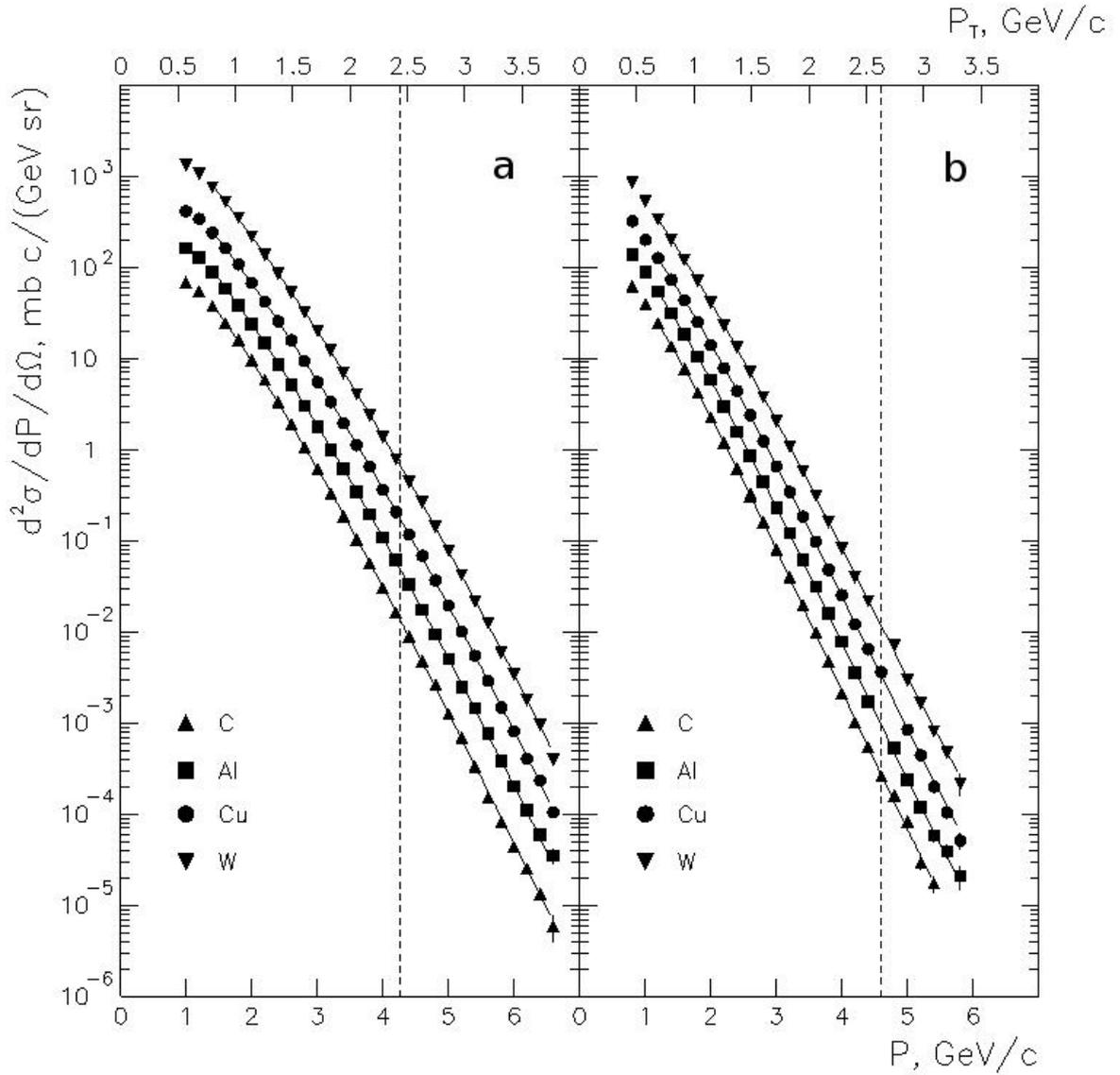

Fig. 2. Differential production cross sections for (a) positively and (b) negatively charged particles as functions of momentum. The upper horizontal axes show the transverse momentum. Vertical dashed lines show the kinematic thresholds for (a) elastic scattering of free nucleons, and (b) single $\pi^-$-meson production in interaction of free nucleons. Curves are plotted for better data perception.

For all four targets used in the experiment the ratio of the yields of positively charged ($h^+$) and negatively charged ($h^-$) particles grows fast with increasing transverse momentum. Fig. 3 shows the $h^+/h^-$ - ratio as a function of $P_T$. The data are shown for this experiment and for comparison the $p/\pi^-$ data obtained at comparable transverse momenta in $pPb$ collisions at 70 GeV proton energy and a registration angle of 9° in the laboratory frame [5] as well as the data [6] obtained in $pW$ interactions at 200 GeV proton energy and a registration angle of 4.4° (in both experiments [5] and [6] an angle of 90° in the center of mass system for free nucleons was chosen).

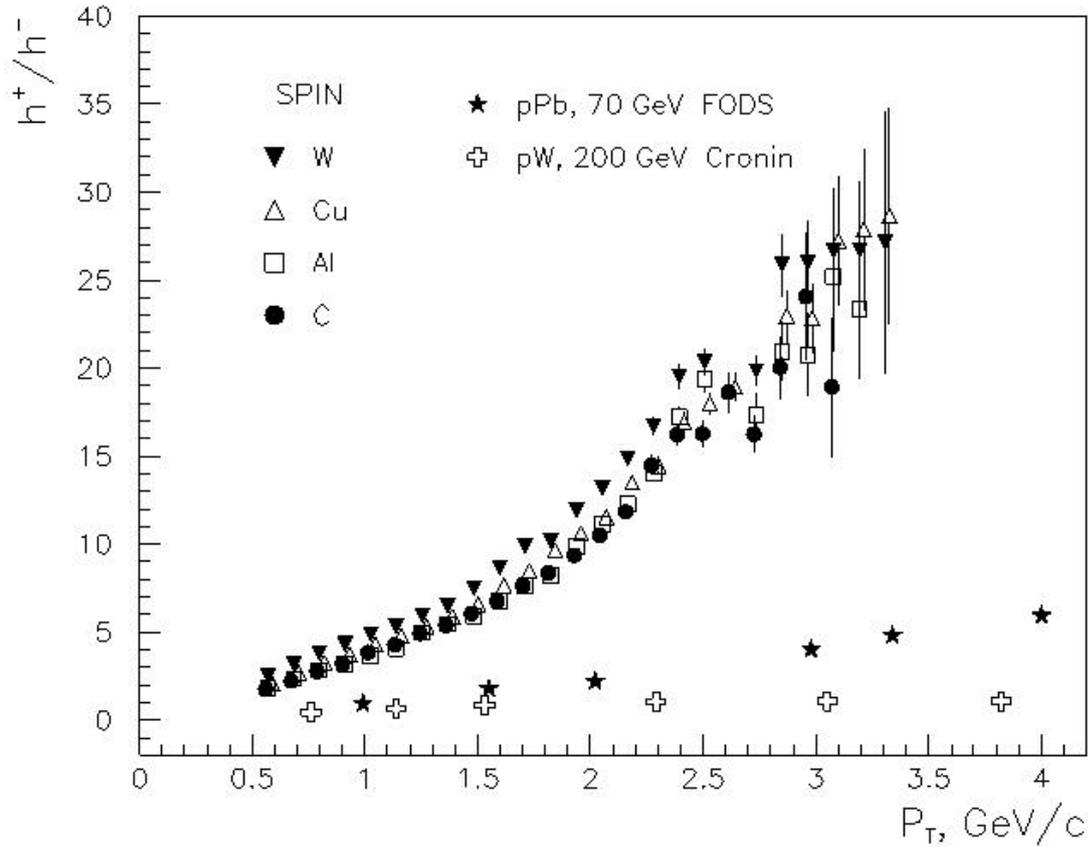

Fig.3. Ratio $h^+/h^-$ data of this experiment for different targets as a function of transverse momentum. The data on ratios $p/\pi^-$, for pPb at 70 GeV [5] and pW at 200 GeV [6] are shown for comparison. The two latter measurements were made at a registration angle of $90^0$ in the center of mass system for free pp -interactions.

It should be noted that the FODS data [5] and the data obtained by Cronin et al. [6] do not go beyond the kinematic threshold of nucleon-nucleon interaction, i.e. they are in the noncumulative domain. It can be seen that in this experiment a much higher $h^+/h^-$ ratio is observed. It should also be noted that with increasing $P_T$ the $h^+/h^-$ data of this experiment become close for all targets.

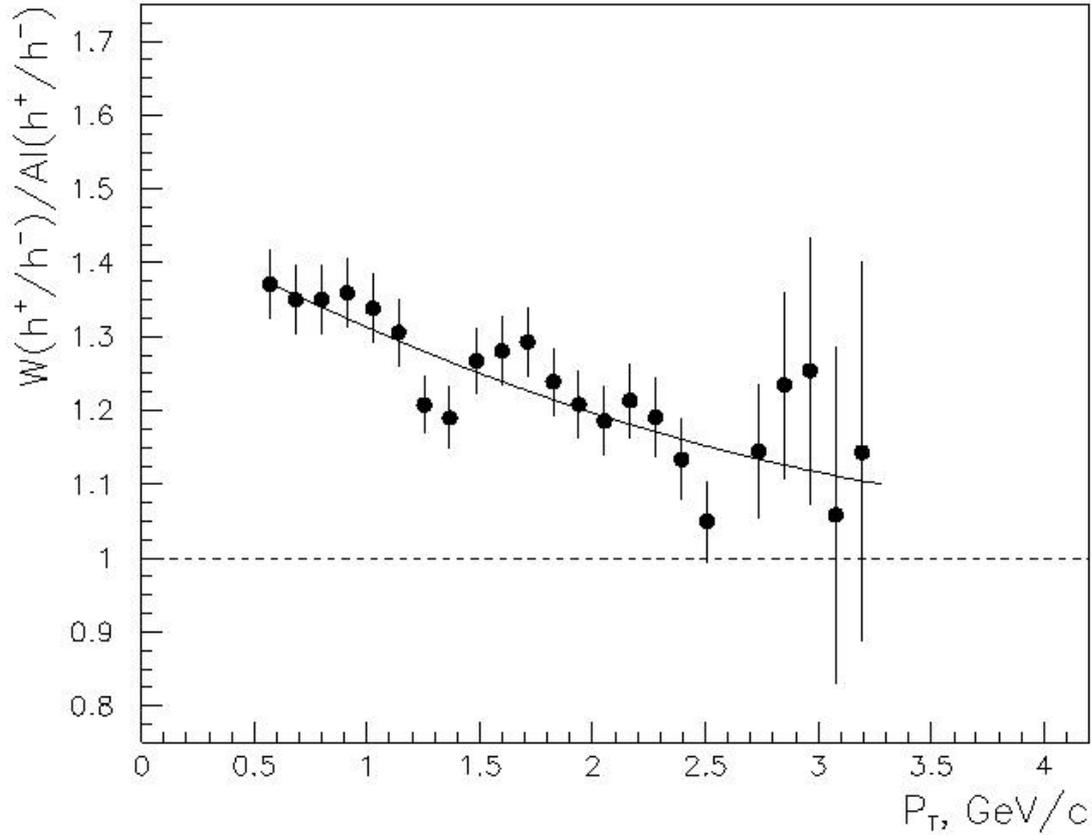

Fig.4. Double ratio of positively and negatively charged particle yields measured on W and Al nuclei. The curve shows a polynomial approximation.

For illustration at Fig. 4 shows the double ratio of $h^+/h^-$ for tungsten and for aluminum. The absence of strong dependence of $h^+/h^-$ on atomic number at high $P_T$ can be interpreted as an indication of a local mechanism of particle production and small contribution of secondary interactions.

Cumulative processes are characterized by high four-momentum transfer. If a cumulative process is considered as a quasi-binary subprocess in which fractions $X_1$ and $X_2$ of four-momenta of the incident particle and the target respectively take part it can be described using these self-similarity variables. This approach is similar to that used in the parton model. The difference is that in this case the squared four-momentum is nonzero, it is equal to the squared fraction of the nucleus (hadron) mass participating in the subprocess. A similarity parameter $s_{min}$ was proposed by V.S. Stavinskiy [7] for description of cumulative processes and high-$P_T$ reactions; this parameter is based on the minimization of a functional comprised of fractions of four-momenta with the meaning of minimum squared sum of fractions of four-momenta of colliding objects necessary for production of the registered pair of particles. This approach makes it possible to determine unambiguously the fractions of four-momenta $X_1$ и $X_2$ corresponding to this minimum for any particular reaction.

The analysis of a large set of inclusive data on particle production in the pre-cumulative and cumulative domains and in inclusive processes at low-$P_T$ subthreshold particle production [8] demonstrated ability to provide qualitative description of invariant particle production cross sections in the form of an incomplete self-similarity solution:

$$f = E\frac{d^3\sigma}{dp^3} = C_1 \cdot A_1^{\frac{1}{3}+\frac{X_1}{3}} \cdot A_2^{\frac{1}{3}+\frac{X_2}{3}} \cdot e^{-\frac{\Pi}{C_2}} \quad (2)$$

where $C_1$ and $C_2$ are constants, same for all reactions, $A_1$ and $A_2$ are the atomic masses of colliding nuclei, $\Pi = \frac{\sqrt{s_{min}}}{2m_N}$ is the similarity parameter where $s_{min}$ is the squared minimum energy for the calculated fractions of four-momenta necessary for producing the observed particle, and $m_N$ is the nucleon mass (in the case of nucleon and nucleus collisions). For proton-nucleus interaction $A_1 = 1$. It was shown in [8] that self-similarity solution (2) provides general description of various processes, especially in A-dependences of inclusive cross sections. According to (2), the ratio of invariant cross sections multiplied by inverse A-dependence should be equal to unity for $pA$ interaction.

$$\frac{f_{(p+A_I)}}{f_{(p+A_{II})}} \times \left(\frac{A_I}{A_{II}}\right)^{-(\frac{1}{3}+\frac{X_2}{3})} = 1 \quad (3)$$

Here $A_I$ and $A_{II}$ are the atomic masses of the target nuclei. It is important to verify whether relation (3) is satisfied for a process similar to a cumulative one.

Fig. 5a shows the ratios of $\pi^-$ production cross sections multiplied by $A^{-(1+X_2)/3}$ for different nuclei. The lower axis of the figure shows the momentum, the upper - quantity $X_2$. It can be seen from this figure that all four ratios W/C, W/Al, W/Cu, and Cu/Al are close to a constant, however, the ratios themselves differ from each other strongly for different pairs of nuclei. The A-dependence of the form $A^{(1+X_2)/3}$ well describes the dynamic dependence of the cross sections on $X_2$, however, it fails to describe a stronger dependence on the mass number of the nucleus observed in this experiment. By using the A-dependence of the form $A^{(\alpha+X_2)/3}$ it was found that the presented negative pion production cross sections correspond to the parameter α=2.45±0.04. This is illustrated in Fig. 5b; this figure shows the ratios of the same cross sections as in Fig. 5a, each multiplied by $A^{-(2.45+X_2)/3}$.

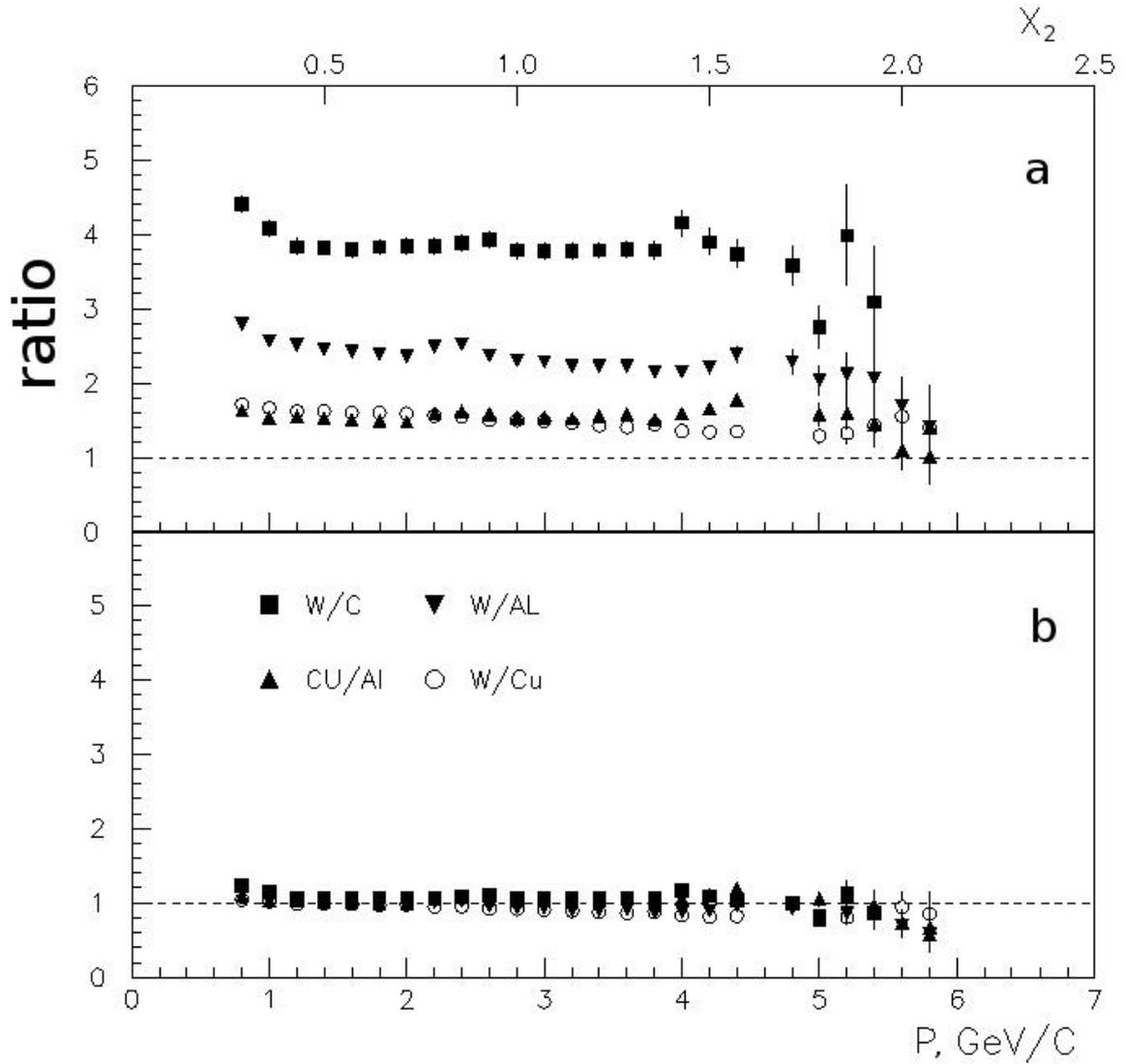

Fig.5 Ratio of cross sections of negative pion production on different nuclei multiplied by inverse A-dependence (see the text). The lower axis shows the momentum, the upper axis, $X_2$. (a) The ratios are obtained using the A-dependence in the form [8] $A^{(1+X_2)/3}$, (b) the ratios are obtained using the A-dependence in the form $A^{(2.45+X_2)/3}$.

## CONCLUSIONS

The momentum spectra of cumulative charged particles in the high-$P_T$ (up to 3.5 GeV/c) domain in the reaction $p + A \rightarrow h + X$ were obtained for the first time. The strong dependence of the measured cross sections on the mass number of the target and the and observed indication for the local character of the particle production processes make it possible to interpret these data as an indication of a large contribution of the processes of interaction of the incident proton with multinucleon (multiquark) configurations of nuclear matter.

The observed strong A-dependence of particle production cross section disagrees at constant part of the exponent with the approach [8] applied for the description of cumulative and sub-threshold reactions at low transverse momenta of registered particles.

The observation of cumulative particles in the high $P_T$ domain makes it possible to conduct correlation studies of the properties of cold super-dense component of nuclear matter. In these correlation studies a cumulative particle can be used as a trigger, and registration of the accompany particles can provide information on the nature of these multinucleon configurations.

We thank the leadership of SRC IHEP for support of this study, the personnel of Accelerator Divisions and Department of Beams for efficient operation of U70 and channel 8. We also thank A.A. Ivanilov for participation in beam runs, assistance in data processing, and fruitful discussions of obtgained results. We also thank A.T. Golovin for invaluable technical support in preparing the setup for operation.